\pdfoutput=1
\documentclass[aps,prd,twocolumn,superscriptaddress,showpacs,preprintnumbers,nofootinbib,longbibliography
]{revtex4-1}

\usepackage{amsmath,amssymb,ifpdf,url,mathrsfs,slashed,multirow,tabularx,type1cm}
\usepackage{graphicx,color}

\usepackage[utf8]{inputenc}
\usepackage{multirow}
\usepackage{collref}
\usepackage{xparse}
\usepackage{subcaption}
\usepackage{hyperref}

\usepackage{geometry}
\usepackage{tikz}
\usetikzlibrary{positioning,arrows,patterns}
\usetikzlibrary{decorations.pathmorphing}
\usetikzlibrary{decorations.markings}
\usetikzlibrary{calc}
\usetikzlibrary{shapes.misc}
\tikzset{
    photon/.style={decorate, decoration={snake}, draw=black, thick},
    fermionnoarrow/.style={draw=black, postaction={decorate}, thick},
    scalar/.style={draw=black, postaction={decorate}, thick, dashed},
    fermion/.style={draw=black, postaction={decorate},decoration={markings,mark=at position .55 with {\arrow{>}}}, thick},
    gluon/.style={decorate, draw=black, decoration={coil,amplitude=4pt, segment length=5pt}, thick},
    vertex/.style={draw,shape=circle,fill=black,minimum size=3pt,inner sep=0pt},
    cross/.style={cross out, draw=black,thick, minimum size=6pt, inner sep=0pt, outer sep=0pt},
    effe/.style={cross out, draw=blue,thick, minimum size=6pt, inner sep=0pt, outer sep=0pt}
}

\begin{document}

\preprint{IPMU16-0065}

\title{TeV-scale Pseudo-Dirac Seesaw Mechanisms \\ in an E$_6$ Inspired Model}

\author{Yi Cai} \email{yi.cai@unimelb.edu.au}
\affiliation{ARC Centre of Excellence for Particle Physics at the Terascale \\
School of Physics, The University of Melbourne, Victoria 3010, Australia}

\author{Jackson D. Clarke}\email{j.clarke5@pgrad.unimelb.edu.au}
\affiliation{ARC Centre of Excellence for Particle Physics at the Terascale \\
School of Physics, The University of Melbourne, Victoria 3010, Australia}

\author{Raymond R. Volkas}\email{raymondv@unimelb.edu.au}
\affiliation{ARC Centre of Excellence for Particle Physics at the Terascale \\
School of Physics, The University of Melbourne, Victoria 3010, Australia}

\author{Tsutomu T. Yanagida}\email{tsutomu.tyanagida@ipmu.jp}
\affiliation{Kavli IPMU (WPI), The University of Tokyo, Kashiwa, Chiba 277-8593, Japan}

\date{\today}

\begin{abstract}
TeV-scale seesaw mechanisms are interesting due to their potential testability at existing collider experiments.  
Herein we propose an E$_6$-inspired model 
allowing a TeV-scale pseudo-Dirac singlet neutrino seesaw mechanism
with naturally sizeable Yukawa couplings of $\mathcal{O}(10^{-2})$.
The model also contains new U(1) gauge interactions (and associated $Z'$ bosons
with which the singlet neutrinos can be produced at colliders),
typically long-lived colour triplet fermions,
SU(2) doublet fermions, and a complex scalar --- potentially all at the TeV-scale.
Additionally we identify three possible explanations for the 750~GeV di-photon excess.  
\end{abstract}

\pacs{
14.60.Pq, 
12.10.Dm 
}

\maketitle

\vspace{-0.2cm}
\section{Introduction}
\vspace{-0.2cm}
The seesaw mechanism \cite{Minkowski1977sc,
Yanagida1979as, GellMann1980vs}
is an elegant way to explain the observed small neutrino masses.
For a right-handed neutrino mass scale $M_N$,
the light neutrino scale is $\sim g^2 v^2/M_N$, 
where $v$ is the vacuum expectation value (vev) of the Higgs boson.
However, the natural scenario $g\sim 1$ 
is impossible to test in present-day experiments due to
the ultra-high right-handed neutrino mass scale.  
A TeV-scale $M_N$ can only be achieved 
at the price of a tiny Yukawa coupling $g\sim \mathcal{O}(10^{-6})$.
The situation is improved (but still far above TeV-scale) 
in the Type~II and Type~III seesaw mechanism variations
(at least in the minimal models)~\cite{ Mohapatra:1980ia, Magg:1980ut,
Schechter:1980gr, Wetterich:1981bx, Lazarides:1980nt, Mohapatra:1980yp,
Cheng:1980qt, Foot:1988aq}.
Alternatively, neutrino masses can be generated at loop-level
in radiative neutrino mass models, first proposed in Ref.~\cite{Zee:1980ai},
where extra loop suppression factors allow sizeable Yukawa couplings.
In this work, we explore a GUT-inspired possibility which realises
a TeV-scale seesaw mechanism with relatively large Yukawa couplings,
and with singlet neutrinos which can be tested at colliders.    

From a grand-unified perspective,
the existence of right-handed neutrinos is motivated
by anomaly cancellation of the U(1)$_{B-L}$ gauge symmetry,
naturally embedded into SO(10).
The right-handed neutrino mass scale then arises from the
spontaneous breakdown of this symmetry.
As we have noted, however, this scenario is eminently untestable.
If we instead embed SO(10) into the E$_6$ gauge group,
a natural candidate for the unified gauge group of string theory 
(see e.g. Ref.~\cite{Gursey:1975ki,Green:1987mn}),
an additional U(1)$_\psi$ may exist.
Anomaly cancellation then requires an extra singlet fermion 
and a ${\bf 5}$+${\bf \bar{5}}$ of the SU(5)~$\subset$~SO(10).
If the additional U(1)$_\psi$ is broken at a very high energy scale, 
then the original seesaw scenario is recovered.
However, the presence of an extra singlet fermion opens up 
a new possibility for the neutrino mass generation:
the right-handed neutrino singlet 
together with the extra singlet
can form a Dirac mass term (which could be TeV scale).
Small neutrino masses can then arise 
purely from the suppression of (multiple powers of) the relatively heavy Dirac mass,
or together with  dimensionally suppressed
couplings of one of the singlets to SM fermions.

This paper is set out as follows.
In Sec.~\ref{sec:model} we describe the model of interest and its low energy matter content 
based on SU$(5) \times$U$(1)_\chi \times$U$(1)_\psi$, 
inspired by the E$_6$ grand unification group.
In Sec.~\ref{sec:numass} we describe the pseudo-Dirac seesaw neutrino mass generation mechanism.
In Sec.~\ref{sec:pheno} and \ref{sec:diphoton} 
we sketch the TeV-scale phenomenology and 
possible explanations for the 750~GeV di-photon excess at the LHC.
We conclude in Sec.~\ref{sec:con}.

\vspace{-0.4cm}
\boldmath  
\section{The Model}\unboldmath
\label{sec:model}
\vspace{-0.2cm}
The high scale symmetry breaking pattern follows 
\begin{align}
{\rm E}_6
\to & \ {\rm SO}(10)\times {\rm U}(1)_\psi \nonumber \\
& \to G_5 \equiv {\rm SU}(5)\times {\rm U}(1)_\chi\times {\rm U}(1)_\psi \; .
\end{align} 
The SM fermions are embedded in a {\bf 27} representation of the E$_6$ group,
which also contains two singlets and {\bf 5}+${\bf \bar{5}}$ of the SU(5) subgroup.    
Two electroweak-scale doublets are embedded in a scalar ${\bf 5}$+${\bf \bar{5}}$
of the SU(5) originally contained also within a {\bf 27} of E$_6$.    
The quantum numbers of the relevant fields are given in Table~\ref{tab:fields}.
The SM Yukawa interactions arise from $\bf 27\, 27\, 27$ in the E$_6$ model,
which decomposes to the subgroup $G_5$ partly as   
\begin{align}
\mathcal{L}^{\rm Yuk} & \supset  g_u \; \psi_{\bf 10}^{\rm SM} \psi_{\bf 10}^{\rm SM} H_{\bf 5} 
           + g_d \; \psi_{\bf 10}^{\rm SM} \psi_{\bf \bar{5}}^{\rm SM} H_{\bf \bar{5}}  \nonumber \\
          & + g_\nu  \; \psi_{\bf \bar{5}}^{\rm SM} N H_{\bf 5} \nonumber \\
          & + g'_1 \; \psi_{\bf \bar{5}}^{\bf 10} N' H_{\bf      5 } 
           + g'_2 \; \psi_{\bf      5 }^{\bf 10} N' H_{\bf \bar{5}} \; ,
\label{eqn:yuk} 
\end{align}
where family indices are omitted. 
All Yukawa couplings are treated as free parameters since 
we only consider an E$_6$-inspired model.
Since the beyond SM states have not been observed,
we postulate that they acquire a sufficiently large mass via 
\begin{equation}
\mathcal{L}^{\rm mass} = 
 \langle \Phi_1 \rangle \psi_{\bf 5}^{\bf 10} \psi_{\bf \bar{5}}^{\bf 10} + 
 \langle \Phi_2 \rangle N N^\prime \; ,
\label{eqn:mass} 
\end{equation}
where $\Phi_1$ arises from the scalar {\bf 27} of E$_6$,
and $\Phi_2$ belongs to a {\bf 351} of E$_6$.
The quantum numbers of both $\Phi_1$ and $\Phi_2$ are included in Table~\ref{tab:fields}.
Apart from the trivial dimension 4 scalar terms, 
the dimension 3 term
\begin{align}
 \mathcal{L}^{\rm D3} = \kappa \Phi_1 H_{\bf 5} H_{\bf \bar{5}}
\end{align}
is also allowed,
which softly breaks an extra global U$(1)$ symmetry. 
At this point, the usual seesaw scenario may already be 
achieved due to a dimension 5 term $(\Phi_1^*)^2 N' N'/M_{\rm Pl}$.
Integrating out the $N'$ gives the right-handed neutrino $N$ a Majorana mass 
of order $\langle \Phi_2 \rangle^2 M_{\rm Pl} / \langle \Phi_1^* \rangle^2$,
which is $\sim 10^{14}$~GeV
if $\langle \Phi_1 \rangle \sim M_{\rm Pl}$ and $\langle \Phi_2 \rangle \sim 10^{16}$~GeV.
In this paper, however, we consider a new low-scale possibility, 
that of $\langle \Phi_1 \rangle \sim$~TeV
(and in some cases $\langle \Phi_2 \rangle \sim$~TeV),
which is of more phenomenological interest. 

There is one last piece of the model before we proceed to discuss phenomenology.
At dimension four (Eq.~\ref{eqn:yuk}),
the coloured components of the $\psi^{{\bf 10}}_{{\bf 5},{\bf \bar{5}}}$
can only decay via the off-shell coloured components of $H_{\bf 5,\bar{5}}$,
whose masses are required to be at least at the GUT scale to avoid fast proton decay.
In the optimistic case of a three-body decay to SM+$N'$,
the decay width
\begin{align}
\Gamma_{\psi_{\bf 5,\bar{5}}^{\bf 10}}^{\rm D4} \lesssim  \frac{1}{32\pi^3} \frac{\langle \Phi_1 \rangle^5}{M_{\rm GUT}^4}
\end{align} 
is $\mathcal{O}(10^{-52})$ GeV for $\langle \Phi_1 \rangle\sim $ TeV.
This is plainly miniscule.
At dimension 5 there exists the term
\begin{align}
\mathcal{L}^{\rm D5}\supset 
\frac{1}{\Lambda }\Phi_1^* \Phi_2^* \psi_{\bf 5}^{\bf 10} \psi_{\bf \bar{5}}^{\rm SM} \; ,
\label{eqn:psid5}
\end{align} 
where $\Lambda$ is some UV suppression scale.
Eq.~\ref{eqn:psid5} induces mixing between $\psi_{\bf 5}^{\bf 10}$ and the SM particles 
and allows two-body decays with 
\begin{align}
\Gamma_{\psi_{{\bf 5}, {\bf \bar{5}}}^{\bf 10} }^{\rm D5} 
 \simeq \frac{\left|f_2 \right|^2}{32\pi} \langle \Phi_1 \rangle \; , 
\end{align}
where $f_2\equiv \langle\Phi_2\rangle/\Lambda$ quantifies the mixing.
We have assumed the vevs of $\Phi_{1,2}$ are real so that  
we can take $\langle \Phi_{1,2}^* \rangle = \langle \Phi_{1,2} \rangle$ for simplicity. 
The lifetime of $\psi^{{\bf 10}}_{{\bf 5},{\bf \bar{5}}}$ must be  
 $\lesssim 1 \,\rm{s}$  to avoid destroying the success of big bang nucleosynthesis (BBN).
This condition is only marginally satisfied with
$\langle \Phi_2\rangle \sim$~TeV and $\Lambda\sim M_{\rm GUT}$.
To relax this restriction,
we can incorporate a scalar field $\Phi_3$
(belonging to a $\bf \overline{78}$ of E$_6$)
into the model,
whose quantum numbers are listed in Table~\ref{tab:fields}. 
This extra scalar opens up additional two-body decay channels for $\psi_{\bf 5,\bar{5}}^{\bf 10}$ via
\begin{align}
\mathcal{L}^{\rm D5} \supset & \
\frac{1}{\Lambda}\left[\Phi_1\Phi_3^* \psi_{\bf 5}^{\bf 10} \psi_{\bf \bar{5}}^{\rm SM} + 
\Phi_3 \psi_{\bf 10}^{\rm SM} \psi_{\bf \bar{5}}^{\bf 10} H_{\bf \bar{5}} \right] ,
\end{align}     
and modifies the total decay width to 
\begin{align}
\Gamma_{\psi_{{\bf 5}, {\bf \bar{5}}}^{\bf 10} }^{\rm D5} \simeq 
\frac{\left|f_2 + f_3\right|^2}{32\pi} \langle \Phi_1 \rangle 
+ \frac{\left|f_3\right|^2}{32\pi} \langle \Phi_1 \rangle, 
\label{eqn:psid5phi3}
\end{align} 
with $f_3\equiv \langle \Phi_3\rangle /\Lambda$.
We have assumed the vev of $\Phi_3$ is real as well. 
\begin{table}[t]
\centering
\begin{tabular}{c|c|ccc}
\hline\hline
                            & E$_6$  & SU(5) & U(1)$_\chi$ & U(1)$_\psi$  \\
\hline
$\psi_{\bf 10}^{\rm SM}$    &  \multirow{6}{*}{27} 	& 10 		& $-1$ 		& 1 \\
$\psi_{\bf \bar{5}}^{\rm SM}$   &                  	& $\bar{5}$ 	& $3$ 		& 1\\
$N$                             &                  	& 1 		& $-5$  		& 1\\
$\psi_{\bf 5}^{\bf 10}$         &                  	& 5 		& $2$ 		& $-2$\\
$\psi_{\bf \bar{5}}^{\bf 10}$   &                  	& $\bar{5}$ 	& $-2$ 		& $-2$\\
$N^\prime$                     &                   	& 1 		& 0 		& 4\\
\hline\hline
$H_{\bf 5}$                    & \multirow{3}{*}{27}  	& 5 		& $2$ 		& $-2$\\ 
$H_{\bf \bar{5}}$              &                      	& $\bar{5}$ 	& $-2$ 		& $-2$\\
$\Phi_1$                     &                        	& 1 		& 0 		&4\\
\hline
$\Phi_2$                    & 351 			& 1 		& $5$ 		& $-5$ \\
\hline
$\Phi_3$                       & $\overline{78}$ 	& 1 		& $5$ 		& 3\\
\hline\hline
\end{tabular}
\caption{The fermionic and scalar matter content
and their quantum numbers under  E$_6$ and $G_5$.
} 
\label{tab:fields}
\end{table}
It is now possible to comfortably satisfy the BBN constraint with 
both $\langle \Phi_1 \rangle$ and $\langle \Phi_2 \rangle$ at the TeV-scale.
Depending on the values of these vevs,
the decay of the TeV-scale coloured $\psi^{{\bf 10}}_{{\bf 5},{\bf \bar{5}}}$ components
can be stable, displaced, or prompt on collider length scales.

To be clear, the presence of $\Phi_3$ is not a necessary condition for the model to be viable.
However we find its inclusion interesting to consider,
as it not only eases the BBN restriction when $\langle\Phi_2\rangle\sim$~TeV, 
but also provides a variety of neutrino mass generation mechanisms
and enriches the possible TeV-scale phenomenology, 
as will be discussed.
We will always consider the vev hierarchy
$\langle\Phi_1\rangle\lesssim \langle\Phi_2\rangle$ when $\Phi_3$ is absent, and
$\langle\Phi_1\rangle\lesssim \langle\Phi_2\rangle \ll \langle\Phi_3\rangle$ otherwise.
Note that the only non-trivial dimension 4 term
involving $\Phi_3$ is $\Phi_1\Phi_1\Phi_2\Phi_3^*$.
This term breaks a global U(1) symmetry under which only the $\Phi_3$ is charged,
and will have implications for the scalar sector.

\vspace{-0.2cm}
\boldmath  
\section{Neutrino Mass Generation}\unboldmath  
\label{sec:numass}
We now proceed to qualitatively discuss the neutrino mass generation in this model.
We work in a simplified scenario with flavour diagonal neutrinos
and it suffices to discuss only one generation of fermions.

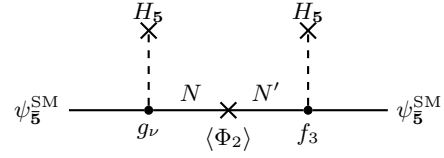
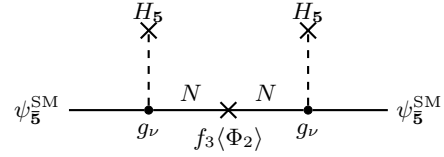
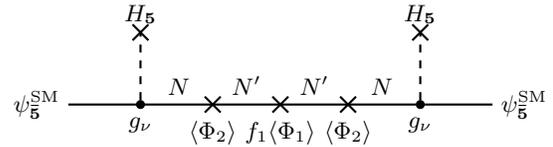
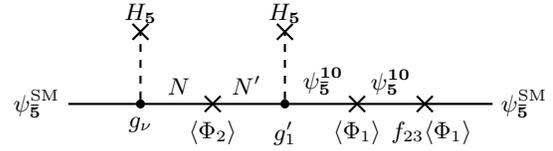
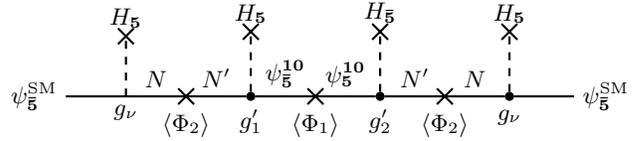
\begin{figure}[t]
\centering
\begin{subfigure}[t]{\linewidth}
\centering
 \begin{tikzpicture}[node distance=1cm and 1cm]
   \coordinate (v0);
   \coordinate[vertex, left = of v0, label=below:$g_\nu$] (v1);
   \coordinate[left = of v1, label=left:$\psi_{\bf {\bar{5}}}^{\rm SM}$] (n1);
   \coordinate[vertex, right = of v0, label=below:$f_3$] (v2);
   \coordinate[right = of v2, label=right:$\psi_{\bf {\bar{5}}}^{\rm SM}$] (n2);
   \coordinate[above = of v1, label=above:$H_{\bf 5}$] (h1);
   \coordinate[above = of v2, label=above:$H_{\bf 5}$] (h2);

   \draw[fermionnoarrow] (n1) -- (v1);
   \draw[fermionnoarrow] (v2) -- (n2);
   \draw[fermionnoarrow] (v1) -- node[label=above:$N$] {} (v0);
   \draw[fermionnoarrow] (v2) -- node[label=above:$N^\prime$] {} (v0);
   \draw[scalar] (v1) -- (h1);
   \draw[scalar] (v2) -- (h2);

   \draw (v0) node[cross, label=below:$\langle \Phi_2 \rangle$] {};
   \draw (h1) node[cross] {};
   \draw (h2) node[cross] {};
 \end{tikzpicture}
 \caption{$\mathcal{O}^{\rm D7}$: linear seeaw}
 \label{fig:numassA}
\end{subfigure}
\vspace{0.2cm}
\begin{subfigure}[t]{\linewidth}
\centering
 \begin{tikzpicture}[node distance=1cm and 1cm]
   \coordinate (v0);
   \coordinate[vertex, left = of v0, label=below:$g_\nu$] (v1);
   \coordinate[left = of v1, label=left:$\psi_{\bf {\bar{5}}}^{\rm SM}$] (n1);
   \coordinate[vertex, right = of v0, label=below:$g_\nu$] (v2);
   \coordinate[right = of v2, label=right:$\psi_{\bf {\bar{5}}}^{\rm SM}$] (n2);
   \coordinate[above = of v1, label=above:$H_{\bf 5}$] (h1);
   \coordinate[above = of v2, label=above:$H_{\bf 5}$] (h2);

   \draw[fermionnoarrow] (n1) -- (v1);
   \draw[fermionnoarrow] (v2) -- (n2);
   \draw[fermionnoarrow] (v1) -- node[label=above:$N$] {} (v0);
   \draw[fermionnoarrow] (v2) -- node[label=above:$N$] {} (v0);
   \draw[scalar] (v1) -- (h1);
   \draw[scalar] (v2) -- (h2);

   \draw (v0) node[cross, label=below:$f_3 \langle \Phi_2 \rangle$] {};
   \draw (h1) node[cross] {};
   \draw (h2) node[cross] {};
 \end{tikzpicture}
 \caption{$\mathcal{O}^{\rm D7}$: inverse seesaw}
  \label{fig:numassB}
\end{subfigure}
\vspace{0.2cm}
\begin{subfigure}[t]{\linewidth}
\centering
 \begin{tikzpicture}[node distance=0.9cm and 0.9cm]
   \coordinate[ label=left:$\psi_{\bf {\bar{5}}}^{\rm SM}$] (n1);
   \coordinate[vertex, right = of n1, label=below:$g_\nu$] (v1);
   \coordinate[ right = of v1] (v2);
   \coordinate[ right = of v2] (v3);
   \coordinate[ right = of v3] (v4);
   \coordinate[vertex, right = of v4, label=below:$g_\nu$] (v5);
   \coordinate[right = of v5, label=right:$\psi_{\bf {\bar{5}}}^{\rm SM}$] (n2);
   \coordinate[above = of v1, label=above:$H_{\bf 5}$] (h1);
   \coordinate[above = of v5, label=above:$H_{\bf 5}$] (h2);

   \draw[fermionnoarrow] (n1) -- (v1);
   \draw[fermionnoarrow] (v1) -- node[label=above:$N$] {} (v2);
   \draw[fermionnoarrow] (v2) -- node[label=above:$N^\prime$] {} (v3);
   \draw[fermionnoarrow] (v3) -- node[label=above:$N^\prime$] {} (v4);
   \draw[fermionnoarrow] (v4) -- node[label=above:$N$] {} (v5);
   \draw[fermionnoarrow] (v5) -- (n2);
   \draw[scalar] (v1) -- (h1);
   \draw[scalar] (v5) -- (h2);

   \draw (v2) node[cross, label=below:$\langle \Phi_2 \rangle$] {};
   \draw (v3) node[cross, label=below:$f_1 \langle \Phi_1 \rangle$] {};
   \draw (v4) node[cross, label=below:$\langle \Phi_2 \rangle$] {};
   \draw (h1) node[cross] {};
   \draw (h2) node[cross] {};
 \end{tikzpicture}
 \caption{$\mathcal{O}^{\rm D9}$: classic seesaw}
  \label{fig:numassC}
\end{subfigure}
\vspace{0.2cm}
\begin{subfigure}[t]{\linewidth}
\centering
 \begin{tikzpicture}[node distance=0.9cm and 0.9cm]
   \coordinate[ label=left:$\psi_{\bf {\bar{5}}}^{\rm SM}$] (n1);
   \coordinate[vertex, right = of n1, label=below:$g_\nu$] (v1);
   \coordinate[ right = of v1] (v2);
   \coordinate[vertex, right = of v2] (v3);
   \coordinate[ right = of v3] (v4);
   \coordinate[ right = of v4] (v5);
   \coordinate[right = of v5, label=right:$\psi_{\bf {\bar{5}}}^{\rm SM}$] (n2);
   \coordinate[above = of v1, label=above:$H_{\bf 5}$] (h1);
   \coordinate[above = of v3, label=above:$H_{\bf 5}$] (h2);

   \draw[fermionnoarrow] (n1) -- (v1);
   \draw[fermionnoarrow] (v1) -- node[label=above:$N$] {} (v2);
   \draw[fermionnoarrow] (v2) -- node[label=above:$N^\prime$] {} (v3);
   \draw[fermionnoarrow] (v3) -- node[label=above:$\psi_{\bf \bar{5}}^{\bf 10}$] {} (v4);
   \draw[fermionnoarrow] (v4) -- node[label=above:$\psi_{\bf 5}^{\bf 10}$] {} (v5);
   \draw[fermionnoarrow] (v5) -- (n2);
   \draw[scalar] (v1) -- (h1);
   \draw[scalar] (v3) -- (h2);

   \draw (v2) node[cross, label=below:$\langle \Phi_2 \rangle$] {};
   \draw (v3) node[ label=below:$g_1^\prime$] {};
   \draw (v4) node[cross, label=below:$\langle \Phi_1 \rangle$] {};
   \draw (v5) node[cross, label=below:$\; \; f_{23} \langle \Phi_1 \rangle $] {};
   \draw (v5) -- (n2);
   \draw (h1) node[cross] {};
   \draw (h2) node[cross] {};
 \end{tikzpicture}
 \caption{$\mathcal{O}^{\rm D9}$: linear seesaw}
  \label{fig:numassD}
\end{subfigure}
\vspace{0.2cm}
\begin{subfigure}[t]{\linewidth}
\centering
 \begin{tikzpicture}[node distance=0.8cm and 0.8cm]
   \coordinate (v5);
   \coordinate[vertex, left = of v5, label=below:$g'_1$] (v4);
   \coordinate[left = of v4] (v3);
   \coordinate[left = of v3, label=below:$g_\nu$] (v2);
   \coordinate[left = of v2, label=left:$\psi_{\bf {\bar{5}}}^{\rm SM}$] (v1);
   \coordinate[vertex, right = of v5, label=below:$g'_2$] (v6);
   \coordinate[right = of v6] (v7);
   \coordinate[vertex, right = of v7, label=below:$g_\nu$] (v8);
   \coordinate[right = of v8, label=right:$\psi_{\bf {\bar{5}}}^{\rm SM}$] (v9);
   \coordinate[above = of v2, label=above:$H_{\bf 5}$] (h1);
   \coordinate[above = of v4, label=above:$H_{\bf 5}$] (h2);
   \coordinate[above = of v6, label=above:$H_{\bf \bar{5}}$] (h3);
   \coordinate[above = of v8, label=above:$H_{\bf 5}$] (h4);

   \draw[fermionnoarrow] (v1) -- (v2);
   \draw[fermionnoarrow] (v2) -- node[label=above:$N$] {} (v3);
   \draw[fermionnoarrow] (v3) -- node[label=above:$N'$] {} (v4);
   \draw[fermionnoarrow] (v4) -- node[label=above:$\psi^{{\bf 10}}_{{\bf \bar{5}}}$] {} (v5);
   \draw[fermionnoarrow] (v5) -- node[label=above:$\psi^{{\bf 10}}_{{\bf 5}}$] {} (v6);
   \draw[fermionnoarrow] (v6) -- node[label=above:$N'$] {} (v7);
   \draw[fermionnoarrow] (v7) -- node[label=above:$N$] {} (v8);
   \draw[fermionnoarrow] (v8) -- (v9);
   \draw[scalar] (v2) -- (h1);
   \draw[scalar] (v4) -- (h2);
   \draw[scalar] (v6) -- (h3);
   \draw[scalar] (v8) -- (h4);

   \draw (v3) node[cross, label=below:$\langle \Phi_2 \rangle$] {};
   \draw (v5) node[cross, label=below:$\langle \Phi_1 \rangle$] {};
   \draw (v7) node[cross, label=below:$\langle \Phi_2 \rangle$] {};
   \draw (h1) node[cross] {};
   \draw (h2) node[cross] {};
   \draw (h3) node[cross] {};
   \draw (h4) node[cross] {};
 \end{tikzpicture}
 \caption{$\mathcal{O}^{\rm D10}$: generalized seesaw}
  \label{fig:numassE}
\end{subfigure}
 \caption{Feynman diagrams contributing to the light neutrino mass. The vevs are assumed real.}
 \label{fig:numass}
\end{figure}

The dimension 5 terms which contribute to the neutrino mass generation are
\begin{align}
\nonumber
-\mathcal{L}^{\rm D5}& \supset
\frac{1}{\Lambda}\left[
\Phi_2\Phi_3 NN + \Phi_1^*\Phi_1^*N^\prime N^\prime + \Phi_2 \Phi_3^* N^\prime N^\prime   \right. \\
& + \Phi_3^* \psi_{\bf \bar{5}}^{\rm SM} N^\prime H_{\bf 5} + (\Phi_1^* \Phi_2^* + \Phi_1 \Phi_3^*)\psi_{\bf 5}^{\bf 10} \psi_{\bf \bar{5}}^{\rm SM}    \nonumber \\
& + \Phi_3 \psi_{\bf \bar{5}}^{\bf 10} N H_{\bf 5} \left. + \Phi_3 \psi_{\bf 5}^{\bf 10} N H_{\bf \bar{5}}  \right] \; .
\label{eqn:massd5} 
\end{align}
Together with Eqs.~\ref{eqn:yuk} and~\ref{eqn:mass}, 
we can write down 
the mass matrix for all the neutral states in the basis of 
$(\nu_L, N, N^\prime, \chi_{\bf \bar{5} }, \chi_{\bf 5})$, with $\chi_{\bf 5, \bar{5}}$ being the neutral components
of $\psi^{{\bf 10}}_{{\bf 5, \bar{5}}}$;
for one generation of fermions,
\begin{align} 
\hspace{-0.4cm}
 \left(
  \begin{array}{ccccc}
   0                          & \cdot & \cdot & \cdot & \cdot \\
   g_\nu \frac{v_u}{\sqrt{2}} & f_3\langle\Phi_2\rangle     & \cdot & \cdot & \cdot \\
   f_3 \frac{v_u}{\sqrt{2}}   & \langle\Phi_2\rangle         & f_1\langle \Phi_1\rangle + f_3\langle\Phi_2\rangle    & \cdot & \cdot \\
   0                         & f_3 \frac{v_u}{\sqrt{2}}      & g_1^\prime\frac{v_u}{\sqrt{2}} & 0      & \cdot\\                
   f_{23}\langle\Phi_1\rangle & f_3\frac{v_d}{\sqrt{2}}   & g_2^\prime\frac{v_d}{\sqrt{2}} & \langle\Phi_1\rangle & 0
  \end{array}
  \right) \; ,
\end{align}
where $f_{23}=f_2+f_3$, $\langle H_{\bf 5,\bar{5}}\rangle = \frac{1}{\sqrt{2}}(0,0,0,0,\pm v_{u,d})$
and only the lower part of the symmetric matrix is shown.
We will assume the vevs of $\Phi_i$ real without loss of generality.
After diagonalising the mass matrix, we have two pairs of pseudo-Dirac fermions of masses 
$\sim \langle \Phi_{1,2}\rangle$ 
and one light neutrino whose mass is expressed 
up to $\mathcal{O}(f_i)$ and $\mathcal{O}(v_i^4)$ as
\begin{align}
m_\nu \simeq & - \frac{f_3 g_\nu v_u^2}{\langle \Phi_2 \rangle}
+\frac{f_3 g_\nu^2 v_u^2}{2\langle \Phi_2 \rangle}
\label{eqn:numass1}\\
&+ \frac{f_1 g_\nu^2 v_u^2 \langle \Phi_1 \rangle}{2 \langle \Phi_2 \rangle^2}
+ \frac{ f_{23} g_1^\prime g_\nu v_u^2}{\langle \Phi_2 \rangle}
\label{eqn:numass2}\\
&-\frac{g_1^\prime g_2^\prime g_\nu^2 v_u^3 v_d}{ 2 \langle \Phi_1\rangle \langle \Phi_2 \rangle^2}.
\label{eqn:numass3}
\end{align}
Each line corresponds to a contribution from the following effective operators,
\begin{align}
&\mathcal{O}^{\rm D7}:  & 
&\psi_{\bf \bar{5}}^{\rm SM} \psi_{\bf \bar{5}}^{\rm SM} H_{\bf 5} H_{\bf 5} \Phi_2^* \Phi_3^*   \; ,\\
\nonumber
&\mathcal{O}^{\rm D9}: &
&\psi_{\bf \bar{5}}^{\rm SM} \psi_{\bf \bar{5}}^{\rm SM} H_{\bf 5} H_{\bf 5} \Phi_1^* \Phi_1^* \Phi_2^* \Phi_2^*   \\
& & &\psi_{\bf \bar{5}}^{\rm SM} \psi_{\bf \bar{5}}^{\rm SM} H_{\bf 5} H_{\bf 5} \Phi_1^* \Phi_1 \Phi_2^* \Phi_3^* \; ,\\
&\mathcal{O}^{\rm D10}: &
&\psi_{\bf \bar{5}}^{\rm SM} \psi_{\bf \bar{5}}^{\rm SM} H_{\bf 5} H_{\bf 5} H_{\bf 5}H_{\bf \bar{5}} \Phi_1^* \Phi_2^* \Phi_2^*  \; ,
\end{align}
suppressed by at most one power of $\Lambda$.
The Feynman diagrams for each contribution are shown in Fig.~\ref{fig:numass}.
Clearly, even without $\Phi_3$
an appropriate neutrino mass can still be achieved by Eq.~\ref{eqn:numass3} (Fig.~\ref{fig:numassE}) alone
with $g_\nu^2 g_1^\prime g_2^\prime \sim (10^{-2})^4$ 
and $\langle \Phi_2 \rangle \sim $ TeV.
This contribution can be interpreted as a generalised seesaw mechanism.
If Eq.~\ref{eqn:numass3} is not dominant, 
we arrive at various interesting seesaw mechanisms.     
The two terms in Eq.~\ref{eqn:numass1} arise from Figs.~\ref{fig:numassA} and \ref{fig:numassB},
and have an origin in the same spirit as those of the linear~\cite{Barr:2003nn} 
and inverse~\cite{Mohapatra:1986bd, Mohapatra:1986aw} seesaw mechanisms, respectively. 
The first term in Eq.~\ref{eqn:numass2}, shown in Fig.~\ref{fig:numassC}, represents the usual seesaw mechanism
that is dominant for $\langle\Phi_1\rangle\gtrsim M_{\rm GUT}$ and suppressed for $\langle \Phi_1 \rangle \sim$ TeV.  
The second term in Eq.~\ref{eqn:numass2}, shown in Fig.~\ref{fig:numassD}, is yet another linear seesaw mechanism,
receiving contributions from both dimension-9 operators.

\vspace{-0.4cm}
\boldmath  
\section{Phenomenology}\unboldmath  
\label{sec:pheno}

This model provides rich phenomenology in various aspects.
In the next two sections we sketch this phenomenology,
saving a detailed study for future work.

The vector-like fermion fields $\psi_{\bf 5,\bar{5}}^{\bf 10}$
contain an SU(2) doublet and a down-type colour triplet
which acquire masses from $\langle \Phi_1\rangle \sim$~TeV.
We will call these $L$ and $D$, respectively.
They are testable at the LHC.
The vector-like lepton doublets are dominantly pair-produced 
via the Drell-Yan process and can be searched 
for at the LHC in multi-lepton final states~\cite{Falkowski:2013jya}.
The limits heavily depend on their mixing with SM charged leptons.
Assuming mixing only with the muon (tau), 
the limit on the mass of the vector-like lepton
is $m_L \gtrsim 300 \ (275)$ GeV~\cite{Dermisek:2014qca,Kumar:2015tna}.      
As discussed in Sec.~\ref{sec:model},
the vector-like quarks can be relatively long-lived.
For $\langle\Phi_1\rangle\sim$~TeV,
the phenomenologically distinct possibilities are
\begin{align}
 \sqrt{|f_2+f_3|^2+|f_2|^2} \gtrsim 
 \begin{cases}
  3\times 10^{-13} & \text{for } \tau_D \lesssim 1\text{ s} \\
  2\times 10^{-10} & \text{for } c\tau_D \lesssim 10\text{ m} \\
  2\times 10^{-7} & \text{for } c\tau_D \lesssim 1\text{ mm} \\
 \end{cases} ,
\end{align}
defining collider stable, displaced, and prompt decays at the LHC, respectively. 
They will be predominantly pair-produced via gluon fusion,
and will either escape the detector 
leaving charged tracks with large ionisation energy loss if collider stable,
or be stopped in the detector and decay out of time  in the case of low transverse momentum.
Both have been searched for at the LHC~\cite{ATLAS:2014fka, Aad:2013gva, Chatrchyan:2013oca, Khachatryan:2015jha}
and the most stringent bound is set by the former,
requiring $m_D \gtrsim 845$~GeV if $c\tau_D \gtrsim 10$~m.
In the displaced regime, searches for displaced jets \cite{CMS:2014wda,Aad:2015rba} are applicable, 
and the lower limit on $m_D$ 
is roughly 800--1000~GeV for 1~mm~$\lesssim c\tau_D \lesssim 10$~m
\cite{Barnard:2015rba,Csaki2015uza}.
In the case of prompt decay to third generation quarks,
i.e. $D\to Wt$, $Zb$, and $Hb$, 
the bound is $m_D \gtrsim 580$--730~GeV \cite{CMS:2012hfa}.

There are potentially testable 
$Z^\prime$ bosons in the model~\cite{London:1986dk, Langacker:2008yv}.  
If $\Phi_3$ is absent, 
$\langle \Phi_{1,2}\rangle$ will break the two U(1) symmetries
and both $Z^\prime$ bosons can acquire masses near the TeV-scale.  
When $\Phi_3$ is present, 
only one of the $Z^\prime$ bosons is at the TeV-scale;
it is equivalent to $Z^\prime_\eta$ appearing in the literature.
These heavy $Z'$ bosons can be searched for 
in events with di-lepton final states at the LHC, 
and  the current best limit is $m_{Z^\prime}\gtrsim 2.8-3.1$ TeV
for $g^\prime \simeq \sqrt{\frac{5}{3}} g \tan\theta_W $
in the absence of exotic decay modes \cite{ATLASZPrime}.
Depending on the mass spectrum, 
the $Z'$ bosons can also decay into pairs of pseudo-Dirac singlet neutrinos, 
or pairs of $\psi^{{\bf 10}}_{{\bf 5},{\bf \bar{5}}}$ states.

The admixture of $N'$ in the $\nu_L$ state is
$V_{l N^\prime} \simeq g_\nu v_u/\langle \Phi_2 \rangle$,
which can be experimentally probed at colliders 
and in lepton-number violating processes
if $g_\nu$ is sufficiently large \cite{Atre:2009rg}.
Specifically, for a heavy neutrino mass $\lesssim 300$ GeV,
8 TeV LHC searches set an upper limit of $g_\nu\lesssim 1.0$ 
for mixing purely with $\nu_e$ \cite{Aad:2015xaa,Khachatryan:2016olu}. 
Sensitivity is expected to improve by an order of magnitude at LHC Run~2~\cite{Deppisch:2015qwa}.  
Besides this classical electroweak search, 
$N$ and $N^\prime$ can also be pair produced 
via $s$-channel Drell-Yan-like processes 
featuring the heavy $Z'$ gauge bosons,
whose production cross section is not suppressed by any small mixing.


As for leptogenesis, 
the lepton asymmetry is washed out by the process
$N^\prime N^\prime \to  H_{\bf 5} H_{\bf \bar{5}}$ 
through $t$-channel $\psi_{\bf 5,\bar{5}}^{\bf 10}$ 
exchange (see Fig.~\ref{fig:numassE}) with a rate 
\begin{align}
\Gamma \sim {g_1^\prime}^2 {g_2^\prime}^2 \frac{ T^3}{ \langle \Phi_1 \rangle^2} \gg H\sim \frac{T^2}{M_{Pl}} \;
\end{align}
at $T\sim \langle\Phi_2\rangle$,
where $H$ is the Hubble constant and $T$ is the temperature. 
Given the TeV-scale scalar sector, however,
electroweak baryogenesis is a possibility.

We now turn to the scalar sector.
For simplicity we consider a scalar sector 
with only $H_{\bf 5,\bar{5}}$ and $\Phi_1$,
i.e. with both $\Phi_2$ and $\Phi_3$ (if it is present) 
heavy and thus decoupled at low energy.
The potential for this two-Higgs-doublet plus singlet model is 
\begin{align}
\nonumber
\mathcal{V}&= m_{11}^2 H_{\bf 5}^\dagger H_{\bf 5} + m_{22}^2 H_{\bf \bar{5}}^\dagger H_{\bf \bar{5}} + m_\Phi^2 \Phi_1^* \Phi_1 \\
\nonumber
& + \frac{\lambda_1}{2} \left(H_{\bf 5}^\dagger H_{\bf 5}\right)^2
                     + \frac{\lambda_2}{2} \left(H_{\bf \bar{5}}^\dagger H_{\bf \bar{5}}\right)^2  
                    + \frac{\lambda_\Phi}{2} \left(\Phi_1^*\Phi_1\right)^2\\
\nonumber
& + \lambda_3 \; \left(H_{\bf 5}^\dagger H_{\bf 5} \right)\left( H_{\bf \bar{5}}^\dagger H_{\bf \bar{5}}\right) 
  + \lambda_4 \; \left( H_{\bf 5}^\dagger H_{\bf \bar{5}}^\dagger\right)\left( H_{\bf \bar{5}} H_{\bf 5}\right)\\ 
\nonumber
& + \left(\lambda_{1\Phi} \; H_{\bf 5}^\dagger H_{\bf 5} 
  + \lambda_{2\Phi} \; H_{\bf \bar{5}}^\dagger H_{\bf \bar{5}} \right) \Phi_1^* \Phi_1\\
& - \left( \kappa  \; H_{\bf 5}H_{\bf \bar{5}} \Phi_1 + \frac{1}{2}\xi^2 \; \Phi_1\Phi_1 + h.c. \right) \; , 
\end{align}
where the $\xi^2$ term only appears when $\Phi_3$ is present 
(from the dimension 4 term $\Phi_1\Phi_1\Phi_2\Phi_3^*$).
Note that $\kappa\to 0$ and $\xi^2\to0$ are each technically natural limits
associated with the reinstatement of a global U(1) symmetry,
thus each may be naturally small.
Apart from the singlet scalar and its mixing with the doublets, 
the rest of the potential is actually 
the well-studied Type II 2HDM which is subject to limits from
$B \to X_s \gamma$, $H/A\to \tau^\pm\tau^\mp$, 
and the measured Higgs boson properties.
These constraints put the model ``naturally'' in the following setup~\cite{Clarke:2015bea}:
$m_{11}^2 <0$, $m_{22}^2>0$, and $m_{\Phi}^2 <0$,
with $m_{22}\gtrsim 480$~GeV to satisfy the $B \to X_s \gamma$ constraint \cite{Amhis2014hma,Misiak2015xwa}.
The vevs of the scalars are approximately
\begin{align}
v_\Phi & \simeq \sqrt{\frac{2\; (-m_\Phi^2 + \xi^2)}{\lambda_\Phi}} \; ,\\ 
v_u & \simeq \sqrt{\frac{2}{\lambda_1}\left( 
 - m_{11}^2 - \lambda_{1\Phi}v_\Phi^2 + \sqrt{2}\kappa v_\Phi \right)}  \; ,\\
v_d & \simeq v_u \left( \frac{\sqrt{2}\; \kappa \; v_\Phi}{ 2 \; m_{22}^2 + \lambda_{2\Phi} v_\Phi^2 + 
    (\lambda_3 + \lambda_4) v_u^2} \right)   \; .
\end{align}
To avoid an unnatural (and radiatively unstable) cancellation for $v_u$ 
we require $|-\lambda_{1\Phi}v_\Phi^2+\sqrt{2}\kappa v_\Phi| \lesssim \lambda_1 v_u^2$,
which for $v_\Phi\sim$~TeV is satisfied for $\lambda_{1\Phi}\lesssim 0.1$ and $\kappa\lesssim10$~GeV,
implying a moderate to large $\tan\beta\equiv v_u/v_d\gtrsim 5$.
The mass spectrum in the limit 
$m_{22}^2\gg \lambda_{2\Phi}v_\Phi^2,(\lambda_3+\lambda_4)v_u^2$ 
is then approximately
\begin{align}
m_S^2 		&\simeq -m_\Phi^2 + \xi^2  \; , 
& m_h^2 	&\simeq \lambda_1 v_u^2  \; , \\ 
m_a^2 		&\simeq \xi^2 \; , 
& m_A^2		&\simeq m_{22}^2 \; , \\
m_{H^\pm}^2	&\simeq m_{22}^2 \; ,
& m_H^2 	&\simeq m_{22}^2  \; . 
\end{align}
We note that this setup gives an automatic alignment limit \cite{Clarke:2015hta}
in which $H_{\bf \bar{5}}$ has a small $h$ admixture $\approx 1/\tan\beta$
and $h$ appears SM-like.
To be clear, if $\Phi_3$ is not present,
$\xi^2=0$ and the CP-odd scalar $a$ is eaten 
by one of the $Z^\prime$ bosons.  

\vspace{-0.4cm}
\section{Di-photon Excess}
\label{sec:diphoton}

Inspired by the 750 GeV di-photon excess recently observed at the LHC~\cite{diphoton1,CMS:2015dxe},
we identify here the regions of model parameter space which might explain it.
Again, we only aim to sketch the possible explanations,
with a detailed phenomenological study left for future work.
For an overview of the excess see Refs.~\cite{Franceschini:2015kwy,Franceschini:2016gxv}, 
which we cite for the henceforth use of relevant quantities such as required widths.

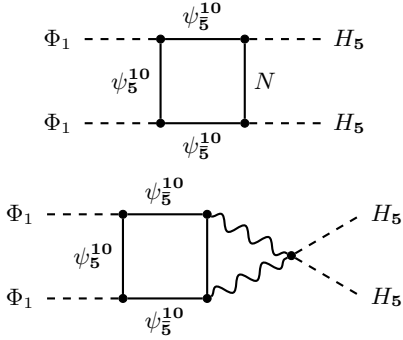
\begin{figure}[t]
\centering
 \begin{tikzpicture}[node distance=1cm and 1cm]
  \coordinate[vertex] (v1); 
  \coordinate[vertex, right = of v1] (v2);
  \coordinate[vertex, below = of v2] (v3);
  \coordinate[vertex, left = of v3] (v4);
  \coordinate[left = of v1, label=left:$\Phi_1$] (s1);
  \coordinate[left = of v4, label=left:$\Phi_1$] (s2);
  \coordinate[right = of v2, label=right:$H_{\bf 5}$] (s3);
  \coordinate[right = of v3, label=right:$H_{\bf 5}$] (s4);

  \draw[fermionnoarrow] (v1) -- node[label=above:$\psi_{\bf \bar{5}}^{\bf 10}$] {} (v2);
  \draw[fermionnoarrow] (v2) -- node[label=right:$N$] {} (v3);
  \draw[fermionnoarrow] (v3) -- node[label=below:$\psi_{\bf \bar{5}}^{\bf 10}$] {} (v4);
  \draw[fermionnoarrow] (v4) -- node[label=left:$\psi_{\bf 5}^{\bf 10}$] {} (v1);
  \draw[scalar] (v1) -- (s1);
  \draw[scalar] (v2) -- (s3);
  \draw[scalar] (v3) -- (s4);
  \draw[scalar] (v4) -- (s2);
 \end{tikzpicture}
 \\
  \begin{tikzpicture}[node distance=1cm and 1cm]
  \coordinate[vertex] (v1); 
  \coordinate[vertex, right = of v1] (v2);
  \coordinate[vertex, below = of v2] (v3);
  \coordinate[vertex, left = of v3] (v4);
  \coordinate[left = of v1, label=left:$\Phi_1$] (s1);
  \coordinate[left = of v4, label=left:$\Phi_1$] (s2);
  \coordinate[vertex, right = of v2, yshift=-0.55cm] (m1);  
  \coordinate[right = of v2, xshift=1cm, label=right:$H_{\bf 5}$] (s3);
  \coordinate[below = of s3, yshift=-0.1cm, label=right:$H_{\bf 5}$] (s4);

  \draw[fermionnoarrow] (v1) -- node[label=above:$\psi_{\bf \bar{5}}^{\bf 10}$] {} (v2);
  \draw[fermionnoarrow] (v2) -- (v3);
  \draw[fermionnoarrow] (v3) -- node[label=below:$\psi_{\bf \bar{5}}^{\bf 10}$] {} (v4);
  \draw[fermionnoarrow] (v4) -- node[label=left:$\psi_{\bf 5}^{\bf 10}$] {} (v1);
  \draw[photon] (v2) -- (m1);
  \draw[photon] (v3) -- (m1);
  \draw[scalar] (v1) -- (s1);
  \draw[scalar] (m1) -- (s3);
  \draw[scalar] (m1) -- (s4);
  \draw[scalar] (v4) -- (s2);
 \end{tikzpicture}
\caption{Notable corrections to $\lambda_{1\Phi}$. Similar diagrams exist for $\lambda_{2\Phi}$.}
\label{fig:lambda1s}
\end{figure}

Either the neutral CP-even scalar $S$ or the CP-odd scalar $a$
can be identified as the resonant state responsible for the excess.
The couplings of $S$ ($a$) with up to three copies of
the vector-like quarks and vector-like lepton doublets
are taken as free parameters $y_D$ and $y_L$ 
($i\gamma^5 y_D$ and $i\gamma^5 y_L$), respectively.\footnote{These 
Yukawa parameters should be unified at the GUT scale as in Eq.~\ref{eqn:mass}.
They are typically split under renormalisation group evolution so that $y_D>y_L$ at low scale.}
The loops of vector-like quarks generate an effective coupling
of $S$ ($a$) to gluons by which it is produced at the LHC via gluon fusion.
To explain the signal (and escape dijet constraints) 
in models where the width to $gg$ dominates requires
$10^{-6}\lesssim \Gamma_{gg}/m_{S(a)} \lesssim 10^{-3}$,
which is achieved for three copies of $D$ with 
$0.2\lesssim y_D\lesssim 6$
($0.1\lesssim y_D\lesssim 4$)
if $m_D = y_D v_\Phi\approx$~TeV.

After production, $S$ or $a$ must decay with the required branching ratio 
to give the di-photon signature.
Before we discuss this decay,
we mention here a few possible decay modes which
are significantly constrained by experiment and must be sufficiently suppressed.
In the CP-even $S$ case,
constraints from $S\to WW$ searches
require the doublet admixture in $S$ to satisfy
$\lambda_{1\Phi,2\Phi} v_{u,d} /(\lambda_\Phi v_\Phi) \lesssim 10^{-2}$,
which implies $\lambda_{1\Phi}\lesssim 0.04$ and $\lambda_{2\Phi}\lesssim 0.04\tan\beta$
for $\lambda_\Phi v_\Phi\approx$~TeV.
As well, decay to $hh$ is constrained as
\begin{align}
 \Gamma(S\to hh) \simeq \frac{\left( \lambda_{1\Phi} v_\Phi \right)^2}{32\pi m_S} \lesssim 20\  \Gamma(S\to \gamma\gamma).
\end{align}
For $\Gamma(S\to \gamma\gamma)/m_S\sim 10^{-6}$ and $v_\Phi\approx$~TeV
this translates to $\lambda_{1\Phi}\lesssim 0.03$.
Conservatively taking the same bound for the $S\to Hh$ decay (if it is allowed)
limits $\kappa\lesssim 50$~GeV, which is already satisfied.
Thus we require $\lambda_{1\Phi,2\Phi}\lesssim 10^{-2}$ 
in order to be phenomenologically safe.
These couplings receive radiative corrections 
from the diagrams in Fig.~\ref{fig:lambda1s}.
Thus they can indeed be as small as $\mathcal{O}(10^{-2})$ 
without too much fine-tuning, even for $y_{D,L}$ Yukawa couplings of $\mathcal{O}(1)$.
Turning to the CP-odd $a$ case,
there is no strong limit from $WW$ or $hh$ on the $\lambda_{1\Phi,2\Phi}$.
The $a$ inherits couplings to the SM fermions
primarily through its $\sqrt{2}$Im$(H_{\bf \bar{5}})$ admixture $\approx (v/v_\Phi)\cot\beta$.
It will then couple to the SM $\tau,b$ ($t$) fermions 
with a strength 
$i\gamma^5y_{\tau,b}^{SM} v/v_\Phi$
($i\gamma^5 y_t^{SM} \cot^2\beta \ v/v_\Phi$).
The decay to $\tau\tau$ is constrained as
\begin{align}
 \Gamma(a\to \tau\tau)\simeq \frac{(y_\tau^{SM} v/v_\Phi)^2 m_a}{8\pi} \lesssim 6\ \Gamma(a\to \gamma\gamma).
\end{align}
For $\Gamma(S\to \gamma\gamma)/m_S\sim 10^{-6}$
this translates to $(v/v_\Phi)^2\lesssim 1.4$, 
which is clearly not a problem.
In principle $a$ can also decay to $Zh$ through mixing,
but this coupling is suppressed due to the automatic alignment limit.
It is now left to reproduce a $\gamma\gamma$ signal from the resonant decay.
We suggest three possible scenarios.

(1) 
If $\Phi_3$ is absent, $a$ will be the Goldstone boson eaten by a TeV-scale 
$Z^\prime \simeq (Z_\psi + Z_\chi)/\sqrt{2}$ with mass $\sim g'\langle\Phi_1\rangle$
and $S$ can potentially be identified with the 750 GeV resonance.
In the absence of additional decay modes,
the signal is fitted with $\Gamma_{\gamma\gamma}/m_S\simeq 10^{-6}$.
It is possible to generate this width via $L$ and $D$ loops.
If $m_D\approx$~TeV and we include three vector-like fermion copies, 
then the minimal values which work are $y_D\approx 2.3$, and $y_L\approx 0.9$,
with $v_\Phi\approx 430$~GeV.
This occurs when $m_L$ is marginally above the 375~GeV pair production threshold.
It is difficult to see how this scenario could be accommodated without having observed the $Z'$.
As well, for $m_S\approx 750$~GeV we must have $\lambda_\Phi\gtrsim 1.7$,
and a Landau pole is of great concern,
especially given we are motivated by grand unification.

(2)
If $\Phi_3$ is present, $\xi^2$ must be non-zero to avoid a massless goldstone mode.
If $\xi^2 \approx (750\text{ GeV})^2$, 
then $a$ could be the 750~GeV resonance.
In fact, for $a$, 
$\Gamma_{\gamma\gamma}/m_a\sim 10^{-6}$
can be minimally achieved for $m_D\approx$~TeV,
$y_D\approx 1.3$, and $y_L\approx 0.5$,
with $v_\Phi\approx 770$~GeV.
These Yukawa values are within the realm of possibility
from a grand-unified perspective.
In addition, if $\xi^2 \gg -m_\Phi^2$,
it is possible that $S$ and $a$ are both present at 750~GeV 
and appear as a single broad resonance.

\begin{figure}[t]
\centering
 \begin{tikzpicture}[node distance=0.8cm and 0.8cm]
  \coordinate[] (v1); 
  \coordinate[vertex, right=of v1,xshift=0.2cm] (v2); 
  \coordinate[ above right=of v2, yshift=-0.1cm] (v3); 
  \coordinate[ below right =of v2, yshift=0.1cm] (v4); 
  \coordinate[ above right=of v3, yshift=-0.6cm, label=right:$\gamma$] (a1); 
  \coordinate[ below right =of v3, yshift=0.6cm, label=right:$\gamma$] (a2); 
  \coordinate[ above right=of v4, yshift=-0.6cm, label=right:$\gamma$] (a3); 
  \coordinate[ below right =of v4, yshift=0.6cm, label=right:$\gamma$] (a4); 
  \coordinate[ above left=of v1,  label=left:$g$ ] (g1); 
  \coordinate[ below left =of v1, label=left:$g$] (g2); 
  \draw[scalar] (v1) -- node[label=above:$S$] {} (v2);
  \draw[scalar] (v2) -- node[label=above:$a$] {} (v3);
  \draw[scalar] (v2) -- node[label=below:$a$] {} (v4);
   \draw[gluon] (g1) -- (v1);
   \draw[gluon] (g2) -- (v1);
   \draw[photon] (v3) -- (a1);
   \draw[photon] (v3) -- (a2);
   \draw[photon] (v4) -- (a3);
   \draw[photon] (v4) -- (a4);
  \draw (v1) node[effe] {};
  \draw (v3) node[effe] {};
  \draw (v4) node[effe] {};
 \end{tikzpicture}
\caption{Feynman diagram for the 750 GeV di-photon excess, where 
the effective interactions between scalars and gauge bosons are denoted with blue crosses. 
}
\label{fig:lhc}
\end{figure}
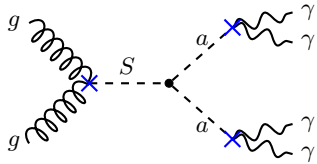

(3)
Instead, it could be that $m_a^2\simeq \xi^2\ll -m_\Phi^2$, say below the GeV-scale.
The $S\to aa$ decay mode,
through the quartic coupling $\lambda_\Phi \left(\Phi_1^*\Phi_1\right)^2$,
will then easily dominate.
The decay width is
\begin{align}
\Gamma_{a a}\simeq \frac{\left(\lambda_\Phi v_\Phi\right)^2}{32\pi m_S}\sim \frac{\lambda_\Phi}{ 16\pi} m_S \; ,
\end{align}
which can imply a broad resonance, of order per cent, if $\lambda_\Phi\sim 1$.  
Subsequently $a$ can decay to a pair of photons 
through loops of $D$, $L$, and SM fermions.
The $a\to \gamma\gamma$ decay mode can be dominant for $m_a<2m_\mu\approx 210$~MeV.
At this point, however, the lifetime is non-negligible, of $\mathcal{O}(1\text{ cm})$.
Still, as long as $a$ decays before the electromagnetic calorimeter,
the 750~GeV di-photon excess is recreated via
$gg\to S\to aa\to (\gamma\gamma)(\gamma\gamma)$, 
with the highly collimated (and displaced) photon pairs 
reconstructed as single prompt photons,
as shown in Fig.~\ref{fig:lhc}.
In this scenario, it is likely that many events are lost due 
to $a$ decaying outside the detector,
and insufficient triggering for displaced events.
Even so, since this final state branching ratio can be $\approx 100\%$
even for a large production cross section via gluon fusion,
a low efficiency can be tolerated.
The displaced $a$ decay to $ee$ will be constrained by
searches for displaced leptonic jets 
appearing at the edge of the electromagnetic calorimeter \cite{Aad:2014yea,Clarke:2015ala},
however it has been noted that this displaced decay channel
has the capacity to itself fake a prompt photon \cite{Agrawal:2015dbf,Tsai:2016lfg}.
Lastly, note that although similarly light pseudoscalars
are excluded by meson decays and fixed target experiments 
unless mixing with the SM fermions is $\lesssim 10^{-4}$ \cite{Dolan:2014ska},
the $\cot^2\beta$ suppression of the $a$ coupling to up-type quarks
(which appear in the loop in meson decays)
opens up gaps in experimental coverage in the region of interest.

\vspace{-0.4cm}
\section{Conclusion}
\label{sec:con}
We have proposed an E$_6$ inspired $G_{SM}\times$U$(1)_\chi\times$U$(1)_\psi$ gauge group model
in which a TeV-scale seesaw mechanism can be naturally realised
without invoking extremely small Yukawa couplings.  
In addition to the SM fermions, 
the fermionic sector consists of (three copies of)
a vector-like down-type quark, 
a vector-like lepton doublet, 
and a pair of SM singlets,
embedded in a $\bf 27$ of E$_6$.
The scalar sector contains two Higgs doublets 
and two or three additional scalar SM singlets $\Phi_{1}$, $\Phi_2$ 
and (optionally) $\Phi_3$,
arising from a $\bf 27$, $\bf 351$ and $\bf \overline{78}$ 
of E$_6$, respectively.
$\Phi_1$ gives mass to the exotic vector-like fermions,
while $\Phi_2$ forms a Dirac mass term with the two singlet fermions.
The two U(1) groups are broken by $\langle\Phi_{1,2,3}\rangle$,
and it is possible that they survive down to the TeV-scale.
The neutrino mass generation manifests in different regions of parameter space 
as various types (or a combination) of seesaw mechanisms including  
a classic seesaw, linear seesaw, and inverse seesaw.
The Yukawa couplings for the neutrinos $g_{\nu}$
can naturally be sizeable, of $\mathcal{O}(10^{-2})$, 
to generate the observed neutrino masses with $\langle \Phi_{1,2}\rangle \sim$~TeV;
then there exists a pseudo-Dirac neutrino pair at the TeV-scale.
Even if $\langle\Phi_{2,3}\rangle \gg $~TeV 
and the pseudo-Dirac neutrino pair decouples,
the scalar modes within $\Phi_1$,
and the vector-like down-type quarks and leptons
are still relevant to TeV-scale physics as long as 
$\langle \Phi_1 \rangle\sim$~TeV.  

We have also sketched the rich TeV-scale phenomenology of the model.
The two additional U(1) gauge symmetries imply extra $Z'$ bosons 
which could lie at the TeV-scale.
These provide an extra channel to pair produce the pseudo-Dirac neutrinos at colliders.
The (typically long-lived) vector-like down-type quarks and the lepton doublets can also be tested.   
Lastly we noted three possible explanations for the 
recently observed 750~GeV di-photon excess:
a weakly coupled model with $gg\to S\to \gamma\gamma$, where $S\simeq \sqrt{2}$Re$(\Phi_1)$;
a weakly coupled model with $gg\to a\to \gamma\gamma$, where $a\simeq \sqrt{2}$Im$(\Phi_1)$;
and the unsuppressed decay of $S$ to a pair very light pseudoscalar states
which subsequently decay to (displaced) collimated di-photons
reconstructed as prompt photons,
$gg\to S \to aa\to (\gamma\gamma)(\gamma\gamma)$.    
 

\vspace{-0.4cm}
\section*{Acknowledgments}
YC, JDC and RRV were supported in part by the Australian Research Council.
TTY was supported by  
JSPS Grants-in-Aid for Scientific Research (No.26287039,26104009,16H02176), 
and World Premier International Research Center Initiative (WPI Initiative), MEXT, Japan.
TTY  thanks the ARC Centre of Excellence for Particle Physics 
at the Terascale in the University of Melbourne for its hospitality, where this work was initiated. 
We thank Joshua Ellis and Michael A. Schmidt for valuable discussion. 
\bibliography{ref}

\end{document}